\RequirePackage{fix-cm}
\documentclass[smallextended]{svjour3}       % onecolumn (second format)
\smartqed  % flush right qed marks, e.g. at end of proof
\usepackage{graphicx}
\begin{document}

\title{Entanglement-assisted codeword stabilized quantum codes with imperfect ebits%\thanks{Grants or other notes
}

\titlerunning{EACWS quantum code with imperfect ebits}        % if too long for running head

\author{Byungkyu Ahn         \and
       Jeonghwan Shin   \and \\
       Jun Heo
}


\institute{Byungkyu Ahn  \at
              School of Electrical Engineering, Korea University, Seoul, Korea \\
              Tel.: +82-2-3290-3779\\
              \email{bk440@korea.ac.kr}           %  \\
           \and
           Jeonghwan Shin \at
              School of Electrical Engineering, Korea University, Seoul, Korea \\
              Tel.: +82-2-3290-4881\\
              \email{jhsh@korea.ac.kr}
               \and
 %          Il-Kwon Sohn \at
 %             School of Electrical Engineering, Korea University, Seoul, Korea \\
 %             Tel.: +82-2-3290-3779\\
 %             \email{d2estiny@korea.ac.kr}
 %              \and
           Jun Heo \at
            School of Electrical Engineering, Korea University, Seoul, Korea \\
              Tel.: +82-2-3290-4824\\
              \email{junheo@korea.ac.kr}
}

\date{Received: date / Accepted: date}

\maketitle

\begin{abstract}

%It has been known that performance of quantum error correcting codes(QECCs) can be improved by error-free entanglemet bits(i.e., ebits) in quantum communication systems.
Quantum error correcting codes (QECCs) in quantum communication systems has been known to exhibit improved performance with the use of error-free entanglement bits (ebits).
%In practical situations, the ebits inevitably suffer from errors and this decreases the error-correcting capability of the code.
In practical situations, ebits inevitably suffer from errors, and as a result, the error-correcting capability of the code is diminished.
Prior studies have proposed two different schemes as a solution.
%To solve this problem, previous research has shown two different schemes.
%One is to use only one QECC to correct errors on Bob's side (i.e., on the receiver) and Alice's side (i.e., on the sender) and the other is to use different QECCs on each side.
One uses only one QECC to correct errors on the receiver's side
(i.e., Bob) and on the sender's side (i.e., Alice). The other uses
different QECCs on each side. In this paper, we present a method
to correct errors on both sides by using single nonadditive
\textbf{Entanglement-assisted codeword stabilized quantum error
correcting code(EACWS QECC)}.
%In this paper, we show how to correct both side errors using the single nonadditive EA-CWS QECC.
We use the property that the number of effective error patterns decreases as much as the number of ebits. This property results in a greater number of logical codewords using the same number of physical qubits.
%We use the property that number of effective error patterns decreases as many as number of ebits. This property results in more number of logical codewords using the same number of physical qubits.
%In EA-CWS code, each stabilizer generator has a single X operator in a different position. Given this property, we find the equivalent transformation relation of the Pauli errors on Bob's side (i.e., on the receiver) and Alice's side (i.e., on the sender) through the operation of the stabilizer generator. Using this relation of equivalency, our scheme simultaneously corrects errors on both Bob and Alice's sides using only one QECC.

 %\and Quantum error correcting code}
\end{abstract}

\section{Introduction}
\label{intro} Over the past two decades research on quantum computing and communications systems has increased.
Quantum error-correcting codes (QECCs) are indispensable to implement practical quantum computing and communication systems since it is not feasible to maintain a quantum state, compute with qubits, or experiment with quantum phenomena without QECCs.
The developments in QECC research have been rapid over the past two
decades as well. The stabilizer formalism \cite{{681315},{1997}}
provides a general framework to construct a QECC as well as an
unified view of quantum and classical-error correcting code. A
classical linear block code with the dual-containing property
\cite{PhysRevA.54.1098} can be converted into a QECC by using stabilizer formalism.

Furthermore, codeword stabilized (CWS) quantum codes
\cite{Cross:2009jo} have also been introduced. CWS quantum code
offers the first unified framework that includes both additive and
non-additive code. It is defined by both a graph
\cite{{012308},{02231}} and classical binary code. Word
stabilizers for the CWS code are generated according to the graph,
and they change any Pauli errors consisting of $X$, $Y(=XZ)$, and
$Z$ operators into effective errors consisting of only the $Z$
operator. By using this feature, any Pauli error can be
transformed into a binary error, with bit 1 for the $Z$ operator
and bit 0 for the $I$ operator.

Entanglement-assisted quantum error correcting code (EAQECC)
\cite{{Brun20102006},{3824-3851},{PhysRevA.76.062313}} is an extended version of standard QECC. EAQECC uses maximally entangled qubits (ebits) shared by the transmitter and receiver. By using these ebits, the EAQECC is not subject to the dual-containing constraint and has a larger minimum distance.

Entanglement-assisted codeword stabilized (EACWS)
quantum codes \cite{PhysRevA.84.062321} has been recently established.
EACWS quantum code can be constructed as nonadditive code of a higher dimension than that of EAQECC with the same number of physical qubits.

Most studies on entanglement-assisted quantum codes have assumed that errors do not occur on the shared ebits from the receiver's side because ebits on the receiver's side do not pass through the transmit channel. However, in practice, receiver-side ebits also suffer from errors, and this reduces the error correcting ability of the code. The following works have taken into account the imperfect ebits.

Shaw et al.  \cite{PhysRevA.78.012337} presented an EAQECC that
corrects errors on both the sender's qubits and the receiver's
shared ebits. They showed for the first time that a Steane code is
equivalent to a [[6,1,3;1]] EAQEC code for correcting a single
error on the receiver's  (i.e. Bob's) ebits. \textbf{Wilde et al.}
\cite{arXiv:1010.1256} simulated entanglement-assisted quantum
turbo codes when the ebits on Bob's side are imperfect. Their aim
was to analyze the effect that ebit noise has on
entanglement-assisted quantum turbo-code performance. Lai and Brun
studied a practical case where errors on the receiver's side can
be corrected. They presented two different schemes
\cite{PhysRevA.86.032319} to correct errors on the receiver's side
and showed an equivalent relationship between $[[n,k,d;c]]$ EAQECC
and $[[n+c,k,d]]$ standard stabilizer code. Based on this
equivalence, EAQECCs can correct errors on the ebits of the
receiver's side. However, when this equivalence does not exist,
the transmitter uses separate EAQECCs to protect the information
qubits while the receiver uses a standard stabilizer code to
protect the ebits.

In this paper, we consider EACWS codes that correct errors on both
sides at the same time. We use the property that the total number
of error patterns decreases through a transition from Pauli errors
to binary errors. \textbf{Transition relation between them is
based on a simple ring graph.} Using this property, we can
generate nonadditive quantum code that has more logical codewords
than additive quantum code with the same number of physical
qubits. In addition, we show that ((6,4,3;1)) EACWS QECC can
correct both side errors even though [[6,2,3;1]] EAQECC does not
have equivalent [[7,2,3]] code.

%To achieve this,we employ the code for representing the equivalence of the errors on Alice's side (i.e. the sender’s side) and those on Bob's side. Using this code, we can find the errors on Bob's side that are equivalent to those on Alice's side by using the appropriate stabilizer generator.  Any Pauli errors can be converted to effective errors with only Z operators, and these can be represented as binary errors. We then search for a set of binary codewords that can correct these binary errors.

The remainder of this paper is organized as follows.
The basics of entanglement assisted codeword stabilized quantum codes are introduced in Section 2. In Section 3, we provide an overview of entanglement-assisted quantum error correcting codes with imperfect ebits.
In Section 4, we describe the proposed scheme for EACWS code with imperfect ebits. We then provide some numerical examples. Finally, we summarize the paper in Section 5.

\section{Entanglement-assisted codeword stabilized (EACWS) quantum code}
\label{sec:1}

%\subsection{Entanglement Assisted - Codeword stabilized (EA-CWS) quantum code}

EACWS code is a class of quantum error correcting code that covers
both additive and nonadditive code. This code is to increase the
capacity of QECCs by using $c$ ebits for CWS quantum codes. An
$((n,K,d;c))$ EACWS quantum code encodes $K$ dimensional code
space into $n$ physical qubits \textbf{with minimum distance $d$}.
In an EACWS code, it is assumed that the receiver's ebits are
error free because the ebits on the receiver's side do not pass
through the channel. We can think of the encoding process for
EACWS codes in the following way.

The initial base state of EACWS code with $n-c$ ancilla qubits and $c$ ebits can be represented by
\begin{equation}
\label{Eq:Initial_unique_state} |S'\rangle=|0\rangle^{\otimes
n-c}|\Phi_+\rangle^{\otimes c},
\end{equation}
where $|\Phi_+\rangle=\frac{1}{\sqrt{2}}(|00\rangle+|11\rangle)$.
The ${c}$ maximally entangled pairs $|\Phi_+\rangle$ are shared
between Alice and Bob.

The set of the word stabilizers $\mathcal{S}'$ for the initial base state that corresponds to the ancilla qubits of $|0\rangle^{\otimes n-c}$  is generated by
\begin{eqnarray}
\left\{\begin{array}{c}
Z_1II\cdots I|I^{\otimes c} \\
IZ_2I\cdots I|I^{\otimes c} \\
\vdots \\
II\cdots IZ_{n-c} I\cdots I |I^{\otimes c} ,\end{array}\right.
\end{eqnarray}
where the operators to the right and the left of $``|"$ respectively act on Alice's and Bob's qubits.

The set of the word stabilizers $\mathcal{S}'_{EA}$  for the initial base state that acts on the ebits is generated by
\begin{eqnarray}
\left\{\begin{array}{c}
I \cdots IZ_{n-c+1}I\cdots I|Z_1 I \cdots I \\
I \cdots IIZ_{n-c+2}I\cdots I|IZ_2 I \cdots I \\
\vdots \\
I\cdots II\cdots IZ_{n}| I \cdots I Z_c,\\
\end{array}\right.
\end{eqnarray}

\begin{eqnarray}
\left\{\begin{array}{c}
I \cdots IX_{n-c+1}I\cdots I|X_1 I \cdots I \\
I \cdots IIX_{n-c+2}I\cdots I|IX_2 I \cdots I \\
\vdots \\
I\cdots II\cdots IX_{n}| I \cdots I X_c .
\end{array}\right.
\end{eqnarray}

For CWS code in a standard form, the initial basis vectors span the code space and are formed by applying the word operators $w_l'$ to the initial base state. Hence, the number of word operators is equal to the dimension of the code space. The initial word operator $\{w'_l\}$  of an EACWS code can be represented by

\begin{equation}
w'_l=X^\mathbf{x_{l}} \otimes Z^\mathbf{v_{l}}X^\mathbf{u_{l}}
|I^{\otimes c},  \textrm{ for}\phantom{1}l=1,\cdots,K ,
\end{equation}
where  $\mathbf{x_{l}}$  is a binary vector of length $n-c$, and
$\mathbf{v_{l}}$ and $\mathbf{u_{l}}$  are binary vectors of
length $c$. The \textit{$X^\mathbf{x_{l}}$}  operators are applied
to $n-c$ ancilla qubits and the
\textit{$Z^\mathbf{v_{l}}X^\mathbf{u_{l}}$} operators are applied
to the $c$ ebits on Alice's side. The identity operator
$I^{\otimes c}$  on the right side means that the word operators
are not applied to Bob's ebits.

The initial basis vectors (i.e., the base states) are given by
\begin{equation}
w'_l|S'\rangle \equiv |w'_l\rangle = X^\mathbf{x_{l}} \otimes
Z^\mathbf{v_{l}}X^\mathbf{u_{l}}|\Phi_+\rangle^{\otimes
c}|0\rangle^{\otimes n-c}.
\end{equation}

The base state does not involve any information qubits. Therefore, we need to encode an information state $|\phi\rangle$  into state  $|\varphi'\rangle$. In this case, the code space is spanned by a linear combination of the states  $|w_l'\rangle$.
%At this point, let us suppose a $K$-dimensional system.
We swap the state $|\phi\rangle$ into the codeword by defining a unitary transformation $U_{w'}$\cite{PhysRevA.84.062321} as follows:
\begin{eqnarray}
U_{w'}(|\phi\rangle\otimes |S'\rangle) &=& |0\rangle \otimes
\sum^{K-1}_{l=0}\alpha_l |w'_l\rangle \nonumber\\
&\equiv& |0\rangle \otimes |\varphi'\rangle.
\end{eqnarray}

One additional step is needed to enable the codewords to correct
errors. A unitary encoding operator $U_E$ is drawn from the
Clifford group and maps the stabilizer generators for the base
state to those of the CWS code in the standard form. By applying
the operator $U_E$, each stabilizer generator has an $X$ operator
on one qubit in a different position and $Z$ operators on qubits
that have relationships in the associated graph. In this paper, we
consider a simple ring graph.

After the unitary encoding process, the word stabilizer is represented as
\begin{equation}\label{eq:Asymptotic-Result2} \left\{\begin{array}{l}
X_1Z_2I \cdots IZ_n|I^{\otimes c} \\
Z_1X_2Z_3I \cdots I|I^{\otimes c} \\
\qquad\qquad \vdots \\
I\cdots I Z_{n-c-1}X_{n-c}Z_{n-c+1}I \cdots I|I^{\otimes c}. \\

\end{array}\right.
\end{equation}
In Equation (\ref{eq:Asymptotic-Result2}), the word stabilizers are generated by encoding them for the initial base state corresponding to the ancilla qubits.

\begin{eqnarray}\label{eq:Asymptotic-Result1} && \left\{\begin{array}{l}
I\cdots I Z_{n-c}X_{n-c+1}Z_{n-c+2}I \cdots I|Z_{1}II\cdots II\\
\qquad\qquad\vdots \\
Z_{1}I\cdots I Z_{n-1}X_{n}|I\cdots I Z_c,\\
\end{array}\right.\\
%\end{equation}
\label{eq:Asymptotic-Result} && \left\{\begin{array}{l}
%\begin{equation} \left\{l}
I\cdots I Z_{n-c+1}I \cdots I|X_1 I\cdots I\\
\qquad\qquad\vdots \\
I\cdots II Z_{n}|I\cdots I X_c.\\
\end{array} \right .
%\end{equation}
\end{eqnarray}
In Equations (\ref{eq:Asymptotic-Result1}) and (\ref{eq:Asymptotic-Result}), the word stabilizers are generated by encoding the word stabilizer of the initial base state corresponds to the ebits.

After applying the unitary encoding operator $U_E$, the base state
$|S'\rangle$ is converted into a state $|S\rangle$:
\begin{equation}
U_E|S'\rangle=|S\rangle.
\end{equation}

Likewise, the word operators are generated by
\begin{equation}
w_l=U_Ew'_lU_E^\dagger.
\end{equation}

\section{Entanglement-assisted quantum error correcting codes with imperfect ebits}

In practical settings, receiver-side ebits also suffer from errors, and this reduces error correcting capability. In this section, we review previous work \cite{PhysRevA.86.032319} that considered two schemes for error correction on the receiver's imperfect ebits.

\label{sec:1}
\subsection{EAQECCs that are equivalent to standard stabilizer codes }

Bowen's [[3,1,3;2]] EAQECC \cite{PhysRevA.66.052313} is equivalent to [[5,1,3]] stabilizer
code, and it can correct an arbitrary single error on both sides. The stabilizer generators of the [[5,1,3]] stabilizer code are
\begin{eqnarray}
\begin{array}{c}
XZZXI, IXZZX \\
XIXZZ, ZXIXZ .\end{array}
\end{eqnarray}

The check matrix for the [[5,1,3]] stabilizer code can be expressed as
follows:
\begin{eqnarray}
\left(%
\begin{array}{ccccccccccc}
  1 & 0 & 0 & 1 & 0 & | & 0 & 1 & 1 & 0 & 0 \\
  0 & 1 & 0 & 0 & 1 & | & 0 & 0 & 1 & 1 & 0 \\
  1 & 0 & 1 & 0 & 0 & | & 0 & 0 & 0 & 1 & 1 \\
  0 & 1 & 0 & 1 & 0 & | & 1 & 0 & 0 & 0 & 1 \\
\end{array}%
\right)
\end{eqnarray}

After row exchange and Gaussian elimination, the check matrix changes into
\begin{eqnarray}
\left(%
\begin{array}{ccccccccccc}
  1 & 0 & 0 & 1 & 0 & | & 0 & 1 & 1 & 0 & 0 \\
  0 & 0 & 1 & 0 & 1 & | & 1 & 1 & 0 & 0 & 0 \\
  0 & 1 & 1 & 0 & 0 & | & 1 & 1 & 1 & 1 & 0 \\
  1 & 1 & 0 & 0 & 0 & | & 1 & 1 & 1 & 0 & 1 \\
\end{array}%
\right)
\end{eqnarray}

The stabilizer generators that correspond to the changed check matrix are
\begin{eqnarray}
\begin{array}{c}
XZZ|XI, ZZX|IX \\
ZYY|ZI, YYZ|IZ. \end{array}
\end{eqnarray}

Based on this result, Theorem 2 in Ref.\cite{PhysRevA.86.032319}
showed that the $[[n-c,k,d;c]]$ EAQECC is equivalent to
$[[n,k,d]]$ standard stabilizer code, and can correct qubit
errors up to $ \lfloor \frac{d-1}{2} \rfloor $ from both sides.

The process for proof is as follows. Assumed that
$[[n,k,d]]$ standard stabilizer code has the set of stabilizer
generators $ \{g_1,g_2,...,g_{n-k} \} $. Then, suppose the check matrix of the stabilizer generators can be expressed by  $[H_X | H_Z]$.
After Gaussian elimination, the check matrix turns into the following form:
\begin{eqnarray}
\left(%
\begin{array}{ccccc}
  A & I_{S \times S} &  | & D & 0  \\
  C & 0 &  | & B & I_{S \times S}  \\
  E & 0 &  | & F & 0  \\
\end{array}%
\right)
\end{eqnarray}
for $ 0\leq S \leq n-k $. Stabilizer generators can be represented
as $g'_1 \bigotimes Z_1$ , ... , $g'_c \bigotimes Z_c$ , $h'_1
\bigotimes X_1$, ... , $h'_c \bigotimes X_c$ , $g'_{c+1}
\bigotimes I$ , ... , $g'_{n-k-c} \bigotimes I$ with
simplified generators $g'_j = UZ_jU^{\dag} , h'_j = UX_jU^{\dag}
(j = 1 , ... , c)$. Therefore, the set of simplified generators is
$\{g'_1 , ... , g'_{n-k-c} , h'_1 , ... , h'_{c}\}$, which indicates $[[n-c,k,d;c]]$ EAQECC.

In addition, they found  some optimal EAQECCs that satisfy the linear
programming bounds and the equivalent relation between $[[n,k,d]]$
standard stabilizer code and $[[n-c,k,d;c]]$ EAQECC as follows:
\begin{eqnarray*}
&&[[15,10,4;5]],[[14,11,3;3]],[[13,9,4;4]],[[13,10,3;3]],[[12,9,3;3]],\\
&&[[11,8,3;3]],[[10,6,4;4]],[[10,7,3;3]],[[9,6,3;3]],[[7,4,3;3]],\\
&&[[8,4,4;4]],[[6,2,4;4]],[[7,3,3,1]],[[6,3,3;2]],[[6,1,5;5]] ,\\
&&[[4,1,3;1]],[[4,1,3,3]],[[3,1,3;2]].\\
\end{eqnarray*}

\label{sec:2}
\subsection{EAQECCs with another quantum code to protect Bob's ebits}

The equivalent relationship is not always satisfied for optimal $[[n-c,k,d;c]]$
EAQECCs and $[[n,k,d]]$ standard stabilizer code. When the equivalence does not exist, it was proposed to use separate QECC in order to protect the ebits. Lai and Brun referred to this scheme as a combination code where the sender uses an
$[[n,k,d_A;c]]$ EAQECC with encoding operator $U_A$ to protect
the information qubits and the receiver uses a separate
$[[m,c,d_B]]$ standard stabilizer code with encoding operator
$U_B$ to protect the ebits. Thus, the entire encoding operator is
represented by $U_A \otimes U_B$, and the notation of the
combination code is $[[n,k,d_A;c]]+[[m,c,d_B]]$.

They also found EAQECCs that are not satisfied by the equivalent relationship between $[[n,k,d;c]]$ EAQECCs and $[[n+c,k,d]]$ standard stabilizer code \cite{{1},{PhysRevA.88.012320}}, and these are:
\begin{eqnarray*}
&&[[n,1,n;n-1]]\ \textrm{for n odd}, [[n,1,n-1;n-1]]\ \textrm{for n even},\\
&&[[5,1,5;4]], [[5,1,4;3]], [[5,1,4;2]], [[5,2,3;2]],\\
&&[[6,1,5;4]], [[6,1,4;3]], [[6,2,4;3]], [[6,2,3;1]],\\
&&[[7,1,5;2]], [[7,1,5;3]], [[7,1,7;6]], [[7,2,5;5]], [[7,3,4;4]], [[7,3,4;3]], [[7,4,3; 2]],\\
&&[[8,1,6;6]], [[8,2,6;6]], [[8,1,6;5]], [[8,3,5;5]], [[8,2,5;4]], [[8,1,4;1]], [[8,3,4;3]], [[8,5,3;2]], \\
&&[[9,1,7;4]], [[9,1,7;5]], [[9,1,7;6]], [[9,1,7;7]], [[9,1,9;8]], [[9,1,7;6]], [[9,1,7;7]], [[9,2,6;6]],\\
&&[[9,1,6;5]], [[9,1,6;6]], [[9,2,5;4]], [[9,5,3;1]],\\
&&[[10,1,8;8]], [[10,1,7;6]], [[10,1,6;5]], [[10,1,6;4]], [[10,2,7;7]], [[10,2,6;5]],\\
&&[[10,2,5;3]], [[10,2,5;2]], [[10,3,6;7]], [[10,3,6;6]], [[10,4,5;5]], [[10,4,5;4]],\\
&&[[13,3,9;10]], [[13,1,11,10]], [[13,1,11;11]],[[13,1,9;8]],\\
&&[[13,1,9;9]], [[15,7,6,8]], [[15,8,6;7]], [[15,9,5;6]].\\
\end{eqnarray*}

A $[[n+m,k,d]]$ standard stabilizer code can correct arbitrary
$\lfloor \frac{d-1}{2} \rfloor $ errors. When compared with
$[[n+m,k,d]]$ standard stabilizer code,
$[[n,k,d_A;c]]+[[m,c,d_B]]$ quantum code uses a smaller number of
qubits going through the noisy channel in order to correct the same number of errors on the transmit
channel. %and offers additional error correcting ability on the receiver's side.

\section{Entanglement-assisted codeword stabilized quantum codes with imperfect ebits}

In this section, we show EACWS code that corrects qubit errors on the transmitter's side and ebit errors on the receiver's side at the same time. Our scheme corrects arbitrary  $\lfloor \frac{d-1}{2} \rfloor$ errors on receiver's side as well as on sender's side. According to the properies of EACWS code, any Pauli error can be turned into a binary error, and we have found binary codewords to correct the binary errors based on exhaustive search. The advantage of this scheme is that it uses only one QECC to correct errors on both sides, regardless of whether the equivalent relation is satisfied.

\label{sec:1}
\subsection{EACWS quantum code with imperfect ebits using the property of stabilizer generators}

Our scheme corrects errors on Bob's side as well as on Alice's
side by using only one QECC. To this end, we use the property of
the EACWS code in such a way that each stabilizer generator $g_i$
(for $i$ = $1$, $\cdots$, $n$) has a single X operator and
multiple Z operators on the qubits corresponding to the
neighboring vertices of the graph. To correct the ebit errors we
need additional word stabilizers $(h_1 , h_2, ... , h_c)$ as well
as the standard word stabilizers $(g_1 , g_2, ... , g_n)$. The
stabilizer generators for the standard form EACWS code consist of
the following:

\begin{eqnarray} \label{eq:Asymptotic-Result4} &&\left\{\begin{array}{l}
g_1 = X_1Z_2I \cdots IZ_n|I^{\otimes c} \\
g_2 =Z_1X_2Z_3I \cdots I|I^{\otimes c} \\
g_3 =IZ_2X_3Z_4I \cdots I|I^{\otimes c} \\
\qquad\qquad \vdots \\
g_{n-c} = I\cdots I Z_{n-c-1}X_{n-c}Z_{n-c+1}I \cdots I|I^{\otimes c} \\
g_{n-c+1} = I\cdots III Z_{n-c}X_{n-c+1}Z_{n-c+2}I \cdots I|Z_{1}I\cdots II \\
\qquad\qquad \vdots \\
g_{n} = Z_{1}I\cdots I Z_{n-1}X_{n}|I\cdots I Z_c ,\\
\end{array}\right.\\
\label{eq:Asymptotic-Result5} &&\left\{\begin{array}{l}
h_{1} = I\cdots I Z_{n-c+1}I \cdots I|X_1 I\cdots I\\
\qquad\qquad\vdots \\
h_{c-2} = I\cdots I Z_{n-2}II|I\cdots I X_{c-2}II\\
h_{c-1} = I\cdots II Z_{n-1}I|I\cdots II X_{c-1}I\\
h_{c} = I\cdots IIIII Z_{n}|I\cdots III X_c ,\\
%I\cdots I Z_mX_{m+1}Z_{m+2}I \cdots I|Z_1 I\cdots I\\
\end{array}\right.
\end{eqnarray}
where Equation (\ref{eq:Asymptotic-Result4}) is derived from
Equations (\ref{eq:Asymptotic-Result2}) and
(\ref{eq:Asymptotic-Result1}). These stabilizer generators
corresponding to a simple ring graph.
 Equation
(\ref{eq:Asymptotic-Result5}) is identical to Equation
(\ref{eq:Asymptotic-Result}).

The stabilizer generator can transform any single Pauli error on both sides into one or more Z errors, and these Z only errors are referred to as effective errors\cite{Cross:2009jo}.
The effective errors are represented as binary errors since the property that turn Z and I operators into 1 and 0. Thus, binary codewords can be found to correct these binary errors. These binary codewords are converted into word operators that formed the basis of the code space. Since the encoding process needs to only be applied to Alice's side, the word operators cannot have Z operators on the qubits in Bob's side, and thus, the stabilizer generators are repeatedly applied to the word operators until all of the Z operators on Bob's side are removed.

%We use the property such that any Pauli errors can be converted to
%effective errors consisting of only Z operators
%\cite{Cross:2009jo}. Let $E_{A}$ denote a single Pauli error on
%the kth qubit on Alice's side. Then, using the stabilizer
%generator in Equation (\ref{eq:Asymptotic-Result4}), the single errors on Alice’s side
%can be transformed into one or more Z errors (consisting
%exclusively of Zs). The X or Y operators in $E_{A}$ can be removed
%using the operation between the stabilizer generators $g_{A}$ and
%$E_{A}$ (for A = 1, $\cdots$, n).

%Any Pauli errors on Bob's side $E_{B}$ (for B = 1, $\cdots$, c)
%can also be converted to effective errors, which only have Z and I
%operators, by multiplying them with the appropriate stabilizer
%elements in Equation (\ref{eq:Asymptotic-Result5}).

%Effective errors are represented as binary errors using the property that turn Z and I operators into 1 and 0. Thus, binary
%codewords can be found to correct these binary errors. These
%binary codewords are converted into word operators, the basis of
%the code space. Because the encoding process need only be applied
%to Alice’s side, the word operators cannot have Z operators on
%the qubits on Bob's side. Stabilizer generators are repeatedly
%applied to the word operators until all of the Z operators  on
%Bob’s side are removed.

\textbf{As we mentioned above, finding EA-CWS code with imperfect ebits is
very similar to it with perfect ebits\cite{PhysRevA.84.062321}.
However, some pairs of Pauli errors on receiver's side and
transmitter's side have the same effective error. In the case of
our scheme with minimum distance of three, the number of these
pairs is the same as the number of ebits as following table 1.
\begin{table}[!h]
\renewcommand{\arraystretch}{1.3}
\caption{In the case of $((n,K,3;c))$ EACWS quantum code, pairs of
errors that have the same effective error.} \centering
\begin{tabular}{|c|c|c|c|}
\hline   & \bfseries  Single  & \bfseries Stabilizer generator  & \bfseries Equivalent  \\
 Number  & \bfseries   X error & \bfseries applies to  & \bfseries  single error\\
 & \bfseries   on Bob's side             & \bfseries    two equivalent errors                & \bfseries on Alice's side  \\
\hline\hline
$1$ & $I \cdots II_n|X_1II \cdots II_c $ & $h_1$ & $Z_1I \cdots I_n|I \cdots I_n$  \\
\hline
$2$ & $I \cdots II_n|IX_2I \cdots II_c $ & $h_2$ & $IZ_1 \cdots I_n|I \cdots I_n$  \\
\hline
$ \vdots $ & $ \vdots $ & $ \vdots $ & $ \vdots $ \\
\hline
$c-1$ & $I \cdots II_n|III \cdots X_{c-1}I_c $ & $h_{c-1}$ & $I \cdots IZ_{c-1}I \cdots  I_n|I \cdots I_n$  \\
\hline
$c$ & $I \cdots II_n|III \cdots IX_c $ & $h_c$ & $I \cdots IIZ_cI \cdots  I_n|I \cdots I_n$  \\
\hline
\end{tabular}
\end{table}
Therefore, the total number of effective errors is
smaller than total number of correctable Pauli errors and it ends
up with higher number of codewords. This is the difference from
EA-CWS code with perfect ebits.}

\textbf{Consider, for instance, the code with $n=7$, $d=3$ and
$c=2$. Suppose the error that occurs on Bob's side, $IIIIIII|XI$.}
We can get an equivalent Pauli error $IIIIIZI|II$ on Alice's side
using the stabilizer generator $h_1=IIIIIZI|XI$. Therefore, two
equivalent errors, $IIIIIII|XI$ and $IIIIIZI|II$, correspond to
the same effective error $IIIIIZI|II$. Due to this reason, the
total number of effective errors is smaller by the number of ebits
than the total number of correctable Pauli errors, resulting in a
higher number of codewords. In $((7,9,3;2))$ EACWS code, we
consider 27 single Pauli errors that consist of 21 errors on the
transmitter's side and 6 errors on the receiver's side. Then, all
Pauli errors are converted into effective errors, including Z and
I operators. In this process, two errors with a single X operator
on the receiver's side have the same effective error with a single
Pauli error on sender's side. \textbf{Due to the presence of two
equivalent error patterns, the total number of effective error
patterns is 25.}

\textbf{In the following section, we consider examples of our
scheme with a minimum distance of three.}

\label{sec:2}
\subsection{Examples of EACWS quantum code with imperfect ebits}

In this section, we provide some examples of the EACWS codes based on our construction. All of the example codes use a base state on a simple ring graph that is identical to a CWS code in standard form. We consider a classical binary-error set and then find classical codes that can correct it through a numerical search, and we then construct the word operators from the set of binary codewords.

\subsubsection{((7,9,3;2)) EACWS code}

A ((7,9,3;2)) code can be constructed from a simple ring graph
with seven vertices by using two ebits with a minimum distance of
three. This nonadditive code has one more dimension of code space
than additive [[9,3,3]] code.

The initial base
state is
\begin{equation}
|S'\rangle = |00000\rangle|\Phi_+\Phi_+\rangle\ .
\end{equation}

The stabilizer generators are generated based on the ring graph as follows:
\begin{eqnarray*}
g_1 &=& XZIIIIZ|II, \\
g_2 &=& ZXZIIII|II, \\
g_3 &=& IZXZIII|II, \\
g_4 &=& IIZXZII|II, \\
g_5 &=& IIIZXZI|II, \\
g_6 &=& IIIIZXZ|ZI, \\
g_7 &=& ZIIIIZX|IZ, \\
h_1 &=& IIIIIZI|XI, \\
h_2 &=& IIIIIIZ|IX. \\
\end{eqnarray*}

%We expect to correct all single errors on both sides, and 27 Pauli error patterns that can be corrected by this code.
%However, two pairs of errors that have the same effective error.
%We can increase capacity of our code since the binary error patterns are reduced by the number of ebits.

%Based on the effective errors, we can derive the binary-error patterns and the binary code that corrects these binary errors. The nine codewords are as follows:
All single errors can be corrected on both sides. Based on the effective errors, nine codewords can be found as follows

\begin{eqnarray*}
&0 0 0 0 0 0 0 | 0 0 ,  1 1 1 0 1 0 1 | 0 1 ,  1 1 1 1 0 0 0 | 0 1 ,  0 0 0 1 0 0 1 | 1 1,  & \\
&0 0 1 0 0 1 0 | 1 1 ,  0 0 1 1 1 1 1 | 1 0 ,  0 1 0 1 1 0 0 | 1 0 ,  0 1 1 1 1 1 0 | 0 1, & \\
&1 1 0 0 0 1 0 | 0 0 . &
\end{eqnarray*}

The word operators are discovered from these binary codewords. The word operators $w'_l$ for the base state $|S'\rangle$ (before applying $U_E$) are

\begin{eqnarray*}
&IIIIIII|II , XXXIXIY|II,  XXXXIIZ|II,  IIIXIZY|II,& \\
&IIXIIYZ|II,  IIXXXYX|II,  IXIXXZI|II,  IXXXXXZ|II,  & \\
&XXIIIXI|II . &
\end{eqnarray*}

and the word operators $w_l$ for this code (after applying $U_E$) are
\begin{eqnarray*}
&IIIIIII|II,  IZZIZZY|II,  IZZZIZX|II,  ZIIZZYX|II,& \\
&ZIZIZXY|II,  IIZZIYI|II,  IZIZIXZ|II,  ZZZZZIX|II,  & \\
&ZZIIIZI|II . &
\end{eqnarray*}

\subsubsection{((9,20,3;1)) EACWS code}

The ((9,20,3;1)) code can also be constructed from a simple ring
graph with nine vertices. This code has two more dimension of code
space than ((10,18,3)) CWS quantum code with a simple ring graph
and the same number of physical qubits.

The initial base state for this code is
\begin{eqnarray}
|S'\rangle = |00000000\rangle|\Phi_+\rangle\ .
\end{eqnarray}

After the encoding operation $U_E$, the stabilizer generators for this code are
\begin{eqnarray*}
g_1 &=& XZIIIIIIZ|I, \\
g_2 &=& ZXZIIIIII|I, \\
g_3 &=& IZXZIIIII|I, \\
g_4 &=& IIZXZIIII|I, \\
g_5 &=& IIIZXZIII|I, \\
g_6 &=& IIIIZXZII|I, \\
g_7 &=& IIIIIZXZI|I, \\
g_8 &=& IIIIIIZXZ|I, \\
g_9 &=& ZIIIIIIZX|Z, \\
h_1 &=& IIIIIIIIZ|X. \\
\end{eqnarray*}

Thirty Pauli error patterns can be corrected with this code. In this case, the number of error pairs that have the same effective error is one, and thus, 30 single Pauli errors can be changed into 29 effective errors (or binary errors), and then, the classical code correcting these effective errors is
\begin{eqnarray*}
&1 1 0 0 0 0 1 0 0 | 1,  1 1 0 0 0 1 0 0 0 | 0,  1 1 0 0 1 0 1 1 1 | 0,  1 1 0 0 1 1 0 1 1 | 1, &  \\
&1 1 1 0 0 0 0 1 0 | 1,  1 1 1 0 1 1 1 0 1 | 1,  1 1 1 1 0 0 0 0 1 | 0,  1 1 1 1 1 1 1 1 0 | 0, &  \\
&0 0 0 0 1 1 1 1 1 | 0,  0 0 0 1 0 0 0 1 1 | 1,  0 0 0 1 1 1 1 0 0 | 1,  0 0 1 1 0 0 1 0 1 | 1, &  \\
&0 0 1 1 0 1 0 0 1 | 0,  0 0 1 1 1 0 1 1 0 | 0,  0 0 1 1 1 1 0 1 0 | 1,  0 1 0 1 0 1 1 0 0 | 0, &  \\
&0 1 0 1 1 0 0 1 1 | 0,  1 0 1 0 0 1 1 0 1 | 0,  1 0 1 0 1 0 0 1 0 | 0,  0 0 0 0 0 0 0 0 0 | 0. &  \\
\end{eqnarray*}

%The following are the word operators $w'_l$ before the encoding constructed from the classical code to correct these errors:
The word operators $w'_l$ for the base state $|S'\rangle$ (before applying $U_E$) are

\begin{eqnarray*}
&XXIIIIXIZ|I,  XXIIIXIII|I,  XXIIXIXXX|I,  XXIIXXIXY|I,  &  \\
&XXXIIIIXZ|I,  XXXIXXXIY|I,  XXXXIIIIX|I,  XXXXXXXXI|I,  &  \\
&IIIIXXXXX|I,  IIIXIIIXY|I,  IIIXXXXIZ|I,  IIXXIIXIY|I,  &  \\
&IIXXIXIIX|I,  IIXXXIXXI|I,  IIXXXXIXZ|I,  IXIXIXXII|I,  &  \\
&IXIXXIIXX|I,  XIXIIXXIX|I,  XIXIXIIXI|I,  IIIIIIIII|I.  &  \\
\end{eqnarray*}

and the word operators $w_l$  for this code (after applying $U_E$) are

\begin{eqnarray*}
&IZIIIIZZX|I,  ZZIIIZIII|I,  ZZIIZIZZZ|I,  IZIIZZIIY|I,  &  \\
&IZZIIIIIX|I,  IZZIZZZZY|I,  ZZZZIIIIZ|I,  ZZZZZZZZI|I,  &  \\
&IIIIZZZZZ|I,  ZIIZIIIIY|I,  ZIIZZZZZX|I,  ZIZZIIZZY|I,  &  \\
&IIZZIZIIZ|I,  IIZZZIZZI|I,  ZIZZZZIIX|I,  IZIZIZZII|I,  &  \\
&IZIZZIIZZ|I,  ZIZIIZZIZ|I,  ZIZIZIIZI|I,  IIIIIIIII|I.  &  \\
\end{eqnarray*}

%\label{sec:3}
\subsubsection{((6,4,3;1))  EACWS code}

According to the Ref.\cite{PhysRevA.86.032319}, a [[6,2,3;1]]
EAQECC not equivalent to standard [[7,2,3]] code. Therefore, when
the sender uses a [[6,2,3;1]] code to protect the information
qubits, the receiver has to use a separate standard stabilizer
code to protect the ebits. On the other hand, our ((6,4,3;1))
EACWS code can simultaneously protect qubits and ebits on both
sides. Based on the simple ring graph, a ((6,4,3;1)) EACWS code
can be generated with six vertices by using one ebits.

The initial base state of this code is
\begin{eqnarray}
|S'\rangle = |00000\rangle|\Phi_+\rangle\ .
\end{eqnarray}

After the encoding operation $U_E$ , the stabilizer generators of this code are
\begin{eqnarray*}
g_1 &=& XZIIIZ|I, \\
g_2 &=& ZXZIII|I, \\
g_3 &=& IZXZII|I, \\
g_4 &=& IIZXZI|I, \\
g_5 &=& IIIZXZ|I, \\
g_6 &=& ZIIIZX|Z, \\
h_1 &=& IIIIIZ|X. \\
\end{eqnarray*}

The total number of single qubit Pauli errors for Alice's and Bob's qubits is 21.
In this case, the total number of binary errors is 20 because two Pauli errors, $IIIIIZ|I$ and $IIIIII|X$, have the same binary error $000001|0$,

The codewords are
\begin{eqnarray*}
& 0 0 0 0 0 0 | 0 , 0 0 1 1 0 0 | 1  , 1 1 0 1 1 1 | 0  , 1 1 1 0 1 1 | 1   & .  \\
\end{eqnarray*}

The word operators before encoding, which is constructed from classical code, are

\begin{eqnarray*}
&IIIIII|I,  IIXXIZ|I,  XXIXXX|I,  XXXIXY|I,  &  \\
\end{eqnarray*}

and the word operators of this code (after applying) are
\begin{eqnarray*}
&IIIIII|I,   ZIZZZX|I,  ZZIZZZ|I,  IZZIIY|I  & . \\
\end{eqnarray*}

\section{Summary}
%In this paper we have presented a scheme to construct a QECC based on the CWS code in standard form. In EA-CWS code with perfect ebits, it is assumed that Bob's ebits are error free. However, our scheme considers an imperfect situation where the receiver's ebits are prone to error. To correct some of the errors on Bob's side, we use a property from the stabilizer generator.
%Using this property, we have shown that the proposed scheme use only one QECC in order to correct errors on both sides.
%Especially, we find a phenomenon ,which the total number of binary errors is diminished by changing process of error patterns from Pauli errors.
%By application of this phenomenon. capacity of QECC can be increased compared to the previous QECCs.
%In cases with a minimum distance of three, our proposed scheme corrects single errors on Alice's or Bob's side. Finally, we constructed several examples of codes $—$ specifically, ((7,9,3;2)) , ((9,20,3;1)) and [[6,2,3;1]] error correcting codes $-$ using the numerical research.
In this paper, we have presented EACWS codes with imperfect ebits.
Based on the simple ring graph, proposed scheme uses only one QECC
to correct errors on both sides. Due to the property that two
different Pauli errors correspond to the same effective error, we
can construct two example codes, a ((7,9,3;2)) and a ((9,20,3;1)),
that have larger codewords than their additive counterparts with
the same number of physical qubits. We also presented a
((6,4,3;1)) EACWS code to protect qubits and ebits on both sides.
\textbf{In the future, we want to find a new code that have better
parameter $K$ by applying a different form of graph. We will also find
another nonadditive EACWS quantum code that have higher minimum
distance.}
 \label{Sec:Conclusions}

%Text with citations \cite{RefB} and \cite{RefJ}.
%as required. Don't forget to give each section
%and subsection a unique label (see Sect.~\ref{sec:1}).

%\paragraph{Paragraph headings} Use paragraph headings as needed.
%\begin{equation}
%\end{equation}

%\begin{figure}
%  \includegraphics{example.eps}
%\end{figure}
%\begin{figure*}
%  \includegraphics[width=0.75\textwidth]{example.eps}
%\end{figure*}
%\begin{table}
%\begin{tabular}{lll}
%\hline\noalign{\smallskip}
%\end{tabular}
%\end{table}

\begin{acknowledgements}
This work was supported by ICT R\&D program of MSIP/IITP. [12-911-04-003, Quantum communication and information processing technology]
\end{acknowledgements}



\end{document}